\documentclass{aa}
\usepackage{psfig,natbib}

\def\chandra{{\it Chandra}}  
\def\xmm{{\it XMM-Newton}}  
\def\asca{{\it ASCA}}  
\def\hst{{\it HST}}  
  
\def\rxte{{\it RXTE}}  
\def\sax{{\it BeppoSAX}}  
  
\def\rosat{{\it ROSAT}}

\def\vla{{\it VLA}}

\def\lum{erg s$^{-1}$}  
\def\flux{erg cm$^{-2}$ s$^{-1}$}  
\def\nh{cm$^{-2}$}  
\def\arcsec{$^{\prime\prime}$}  
\def\deg{$^{\circ}$}  
\def\arcmin{$^{\prime}$}  
\def\ngc{NGC~6251}  
\def\ltsima{$\; \buildrel < \over \sim \;$}  
\def\simlt{\lower.5ex\hbox{\ltsima}} 
\def\gtsima{$\; \buildrel > \over \sim \;$}  
\def\simgt{\lower.5ex\hbox{\gtsima}} 

\begin{document}  
  
\title{The XMM-Newton view of the X-ray halo and jet of NGC~6251}  
  
\normalsize  
  
\author{R. M. Sambruna\inst{1}
\and  M. Gliozzi\inst{1}
\and  D. Donato\inst{1} 
\and  F. Tavecchio\inst{2}
\and  C. C. Cheung\inst{3}
\and  R.F.Mushotzky\inst{4}}
\offprints{rms@physics.gmu.edu}
\institute{George Mason  
University, Dept. of Physics and Astronomy and School of Computational  
Sciences, MS 3F3, 4400 University Drive, Fairfax, VA 22030
\and Osservatorio Astronomico di Brera, via Brera 28, 20121 Milano, Italy
\and Brandeis University, Department of Physics, MS 057, Waltham, MA 02454
\and NASA Goddard Space Flight Center, Code 662, Greenbelt, MD 20771}

\date{Received:21/07/03; accepted:24/10/03}

\abstract{We present an \xmm\ observation of the radio jet and diffuse halo of 
the nearby radio galaxy \ngc. The EPIC spectrum of the galaxy's halo 
is best-fitted by a thermal model with temperature $kT \sim 1.6$ keV 
and sub-solar abundances. Interestingly, an additional hard X-ray 
component is required to fit the EPIC spectra of the halo above 3 keV, 
and is independently confirmed by an archival \chandra\ 
observation. However, its physical origin is not clear.  Contribution 
from a population of undetected Low Mass X-ray Binaries seems 
unlikely. Instead, the hard X-ray component could be due to inverse 
Compton scattering of the CMB photons (IC/CMB) off relativistic 
electrons scattered throughout the halo of the galaxy, or non-thermal 
bremsstrahlung emission. The IC/CMB interpretation, together with 
limits on the diffuse radio emission, implies a very weak magnetic 
field, $B << 1\mu$Gauss, while a non-thermal bremsstrahlung origin 
implies the presence of a large number of very energetic electrons. We 
also detect X-ray emission from the outer ($\sim$ 3.5\arcmin) jet, 
confirming previous \rosat\ findings. Both the EPIC and ACIS spectra 
of the jet are best-fitted by a power law with photon index $\Gamma 
\sim 1.2$, fixed Galactic column density, and 1 keV flux $ 
F_{\rm 1~keV}=2.1$ nJy.  A thermal model is formally ruled out by the 
data. Assuming an origin of the X-rays from the jet via IC/CMB, as 
suggested by energetic arguments, and assuming equipartition implies a 
large Doppler factor ($\delta \sim 10$). Alternatively, weaker beaming 
is possible for magnetic fields several orders of magnitude lower than 
the equipartition field.  
}
\titlerunning{The XMM-Newton view of X-ray halo and jet of \object{NGC~6251}}
\authorrunning{R. M. Sambruna et al.}
\maketitle
   
\section{Introduction}  
\ngc\ is a giant elliptical galaxy at redshift $z$=0.024. It is the  
host of a supermassive black hole with mass $M_{\rm BH}\sim 4-8  
\times10^8 M_{\odot}$ (Ferrarese \& Ford 1999), dynamically estimated  
with \hst.  At radio wavelengths, the source is well studied.  
The radio source measures 1.2\deg\ across the sky with bright hotspots 
that lie toward the edges of its two bright radio lobes (Waggett et al.   
1977). This structure is indicative of many powerful Fanaroff \& Riley 
(1974) type-II radio galaxies. However, its power at 178 MHz (Waggett et 
al. 1977) is below the Fanaroff \& Riley division between FRI/II sources, 
so NGC 6251 appears to be underluminous considering what is expected based 
on its morphology. A prominent radio jet appears to the N-W for 4.5', as 
well as a weak counterjet opposite of the nucleus (Perley et al. 1984). 
The radio jet appears narrow out to $\sim$2', where the 
radio shows a bright knot and more diffuse emission (see Fig. 1). 

Together with NGC~6252, which lies a few arcminutes to the North,  
\ngc\ belongs to the outskirts of the cluster of galaxies Zw  
1609.0+8212 (Young et al. 1979). As discussed in Birkinshaw \& Worrall 
(1993), the cluster does not significantly affect the dynamical 
properties of \ngc.  
  
Previous X-ray imaging studies of \ngc\ with \rosat\ showed the  
presence of an unresolved nuclear source embedded in diffuse thermal  
emission associated with the galaxy's halo (Birkinshaw \& Worrall 1993),  
with no contribution from the cluster gas. X-ray emission from the  
outer radio jet was also detected with \rosat\ (Birkinshaw \& Worrall  
1993), with two compact knots at $\sim$ 4\arcmin\ and 6\arcmin\ from  
the core (Mack et al. 1997a). The latter authors favor a  
thermal origin for the X-rays on the basis of energetic  
arguments. Optical emission from the jet region at $\sim$ 20\arcsec\  
was also claimed (Keel 1988), but never confirmed with \hst.  
  
Comparison of the halo gas pressure to the internal jet pressure led  
Birkinshaw \& Worrall (1993) to conclude that the jet cannot be  
confined by the atmosphere of the galaxy. Furthermore, Mack et  
al. (1997a) proposed that the jet is magnetically confined.  
  
We observed \ngc\ with \xmm\ as part of a project aimed at studying 
the X-ray emission from the various components of the galaxy. With its 
better sensitivity and good resolution, the EPIC camera on-board \xmm\ 
enables a detailed spectroscopic study of the nucleus and extended 
features in the 0.3--10 keV energy range. The analysis of the nuclear 
properties is presented in Gliozzi et al. (2003a; Paper I in the 
following). Here, we concentrate on the X-ray emission from the halo 
and the jet, using the EPIC data and archival \chandra\ and \vla\ 
data.  In this paper, H$_0=75$ km s$^{-1}$ Mpc$^{-1}$ and $q_0=0.5$ 
are adopted. With this choice, 1\arcsec\ corresponds to 446 pc at the 
distance of \ngc\ (97.6 Mpc).  
 
\section{Observations and Data Reduction}   
  
\subsection{XMM-Newton}  
   
\xmm\ carries on-board three sets of CCDs in the focus of the  
telescope, the EPIC pn and the EPIC MOS1 and MOS2 cameras. While the 
pn has a larger effective area and is thus optimized for higher 
signal-to-noise spectroscopy, the MOS offers better imaging 
capabilities. Both the EPIC pn and MOS cameras operate in the energy 
range 0.3--10 keV. 
  
We observed \ngc\ on 2001 March 26 for a duration of $\sim 41$ ks with 
the EPIC pn, and for $\sim 49$ ks with EPIC MOS1 and MOS2. The EPIC pn 
camera was operated with the thin filter, and the MOS cameras with the 
medium filter (to avoid pileup of the bright nucleus; Paper 
I). Details of the reduction procedure were given in Paper I. In 
summary, the data were reprocessed and screened with the 
\xmm\ Science Analysis Software (\verb+SAS+ v.5.3.3) to remove known hot  
pixels and other data flagged as bad; only data with {\tt FLAG=0} were  
used. Investigation of the full--field light curves revealed the  
presence of a period of background flaring at the end of the  
observation.  These events were screened, reducing the effective total  
exposures to $\sim 36$ ks for the EPIC pn and $\sim 43$ ks for the MOS  
cameras.  Background data were extracted from source-free circular  
regions on the same chips containing the source.  There were no signs  
of pile-up in the pn or MOS cameras according to the {\tt SAS} task  
{\tt epatplot}.    
  
The X-ray spectrum of the jet was extracted from an elliptical region 
with semi-axes of 40\arcsec and 12\arcsec, located at $\sim$ 
230\arcsec\ from the nucleus.  The total count rate of the jet in the 
above extraction region in the energy range 0.5-9 keV is $(9.3 \pm 
1.4) \times 10^{-3}$ c/s with EPIC pn, $(2.8 \pm 0.5) \times 10^{-3}$ c/s  
with MOS1, and $(1.5 \pm 0.5) \times 10^{-3}$ c/s with MOS2.   
  
Two different regions were used to investigate the spectral properties 
of the halo: 1) an annular region with radii 35--215\arcsec; 2) a 
series of annular regions, with an increment of the radius of 
30\arcsec, with inner radii ranging from 35\arcsec\ to 215\arcsec.  In 
both cases, the radio jet and the point sources visible in the EPIC 
MOS image were removed. Conservatively, a region of the same size as 
the jet but at the opposite azimuth (where a possible counterjet could 
be present at X-rays) was also excised. The count rate of the halo in 
an extraction annulus with radii 35--215\arcsec\ in the energy range 
0.5--2 keV is 0.37 $\pm$ 0.01 c/s with EPIC pn, 0.102 $\pm$ 0.003 c/s 
with MOS1, and 0.110 $\pm$ 0.003 c/s with MOS2. 
  
Spectral analysis was performed using \verb+XSPEC+ v.11. The EPIC pn  
and MOS spectra were fitted jointly, leaving the relative  
normalizations free to vary to account for inter-calibration  
effects. The data were rebinned with a minimum of 20 counts per bin,  
in order to apply the $\chi^2$ statistics, and fitted using the latest  
calibration files provided by the \xmm\ Guest Observer  
Facility. Improvement of the fit when free parameters were added to  
the fit was evaluated using the F-test, assuming as a threshold for  
improvement $P{\rm _F}$=95\%.  
\begin{figure}  
\psfig{figure=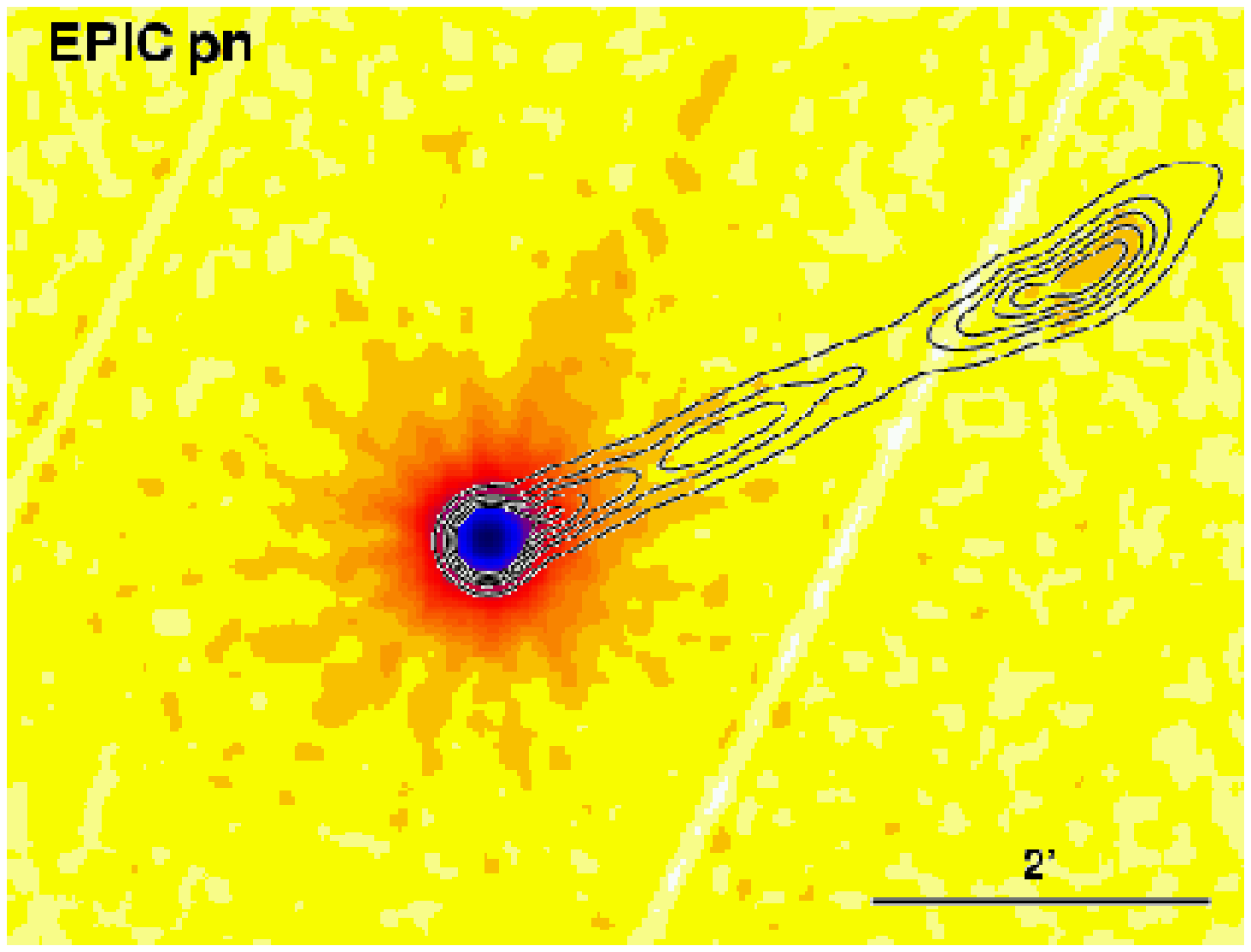,height=7.0cm,width=9.0cm}  
\psfig{figure=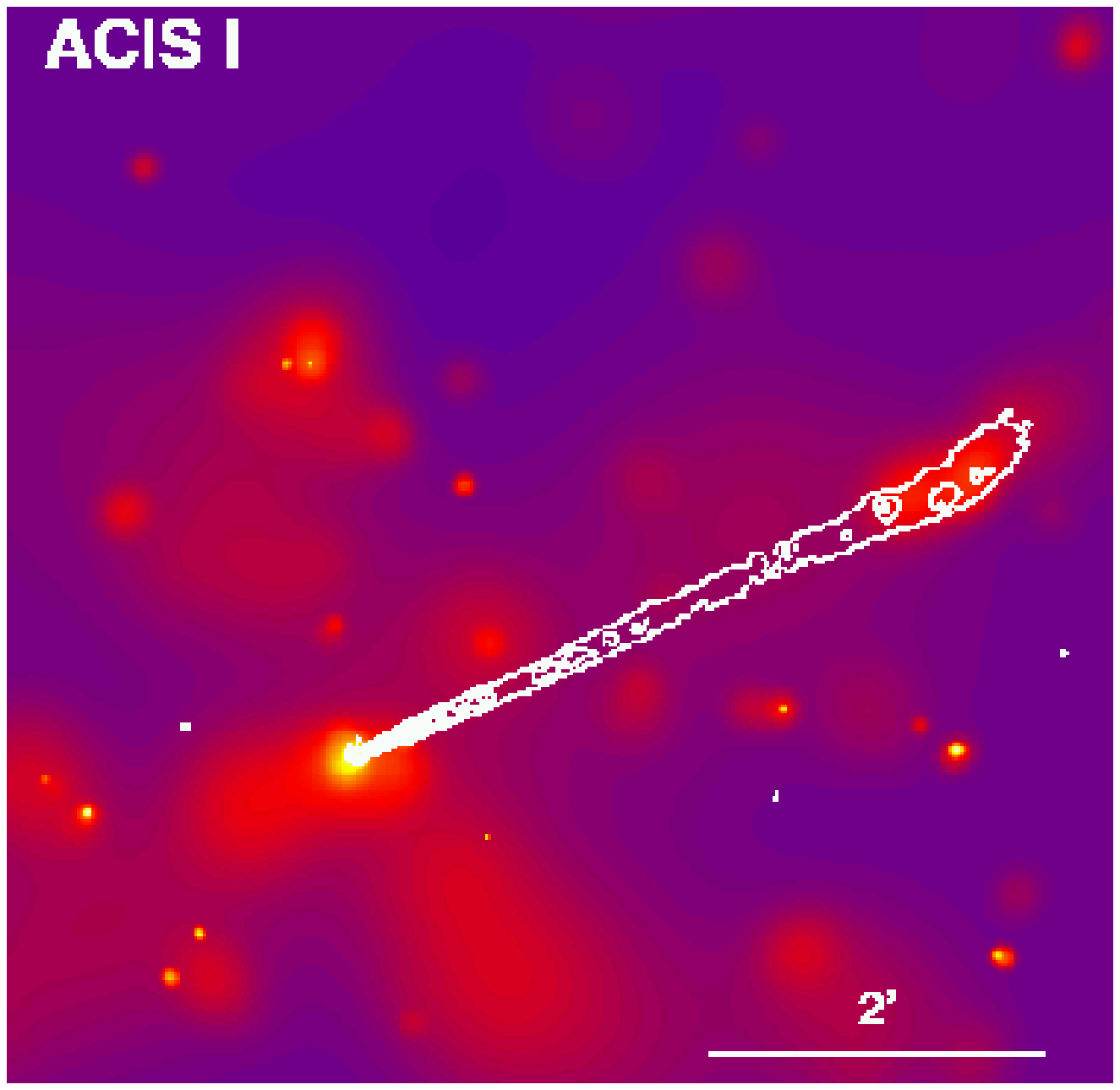,height=7.0cm,width=9.0cm}  
\caption{{\it (a, Top)} EPIC pn image of \ngc\ in 0.3--10 keV and  
{\it (b, Bottom)} \chandra\ ACIS-I image in 0.3--8 keV. The resolution 
of the EPIC image is 10\arcsec, while the resolution of the ACIS image 
is 2\arcsec. In the two panels, the 1.4 GHz \vla\ data are 
overlaid. X-ray emission from the nucleus, diffuse halo, and outer 
jet is apparent.} 
\end{figure}    
\subsection{Chandra}  
  
A 25.4 ks archival \chandra\ image of \ngc\ was investigated to derive  
better spatial constraints on the jet and halo. \chandra\ observed  
\ngc\ on Sept 11, 2000 for 25.4 ks with ACIS-I (PI Kerp). In order to study the  
X-ray emission from the jet, the source was positioned with the  
nucleus on the gap between chips I1 and I2. While this strategy  
prevents a detailed analysis of the nuclear properties, the  
\chandra\ data in principle can be useful to investigate the spatial properties   
of the X-ray halo and jet.   
  
We reduced the ACIS data using standard procedures with the  
\verb+CIAO+2.3  software. We restricted our analysis  
to the 0.3--8.0 keV energy range, where the instrument is better 
calibrated and the background is negligible. We inspected the light 
curves of the background to search for possible background 
fluctuations. No flares have been detected.  Before extracting the 
radial profiles of the sources, we removed field sources from the 
images (see Appendix).  To find the field sources we used the tool 
{\tt wavdetect} with the default value ($10^{-9}$) for the threshold 
for identifying a pixel as belonging to a source. 
  
Using the tools {\tt dmextract} and {\tt dmtcalc}, we extracted the 
radial profiles on a scale of 4\arcmin.  The jet and halo spectra were 
extracted in the same regions as for EPIC (see above), removing the 
point sources and, in the case of the halo, the jet and possible 
counterjet regions. In order to evaluate the instrumental response to 
a point source, we used the method explained in Donato et 
al. (2003). Although a script that automatically performs the same 
task (\verb+ChaRT+) has been released by the \chandra\ X--ray Center, 
this script failed in the creation of the instrumental PSF, since the 
source is positioned in the gap between chips (photons are created but 
not detected). 
  
\subsection{Radio}  
  
In order to explore the spatial coincidence of X-ray features with the 
known radio jet, we obtained data from the archive of the 
NRAO\footnote{The National Radio Astronomy Observatory is a facility 
of the National Science Foundation operated under cooperative 
agreement by Associated Universities, Inc.} {\it Very Large Array} 
(Thompson et al. 1980). We chose a L-band observation from August 15, 
1995 which used the full \vla\ to collect over 3 hours of data 
``on-source'' in its highest resolution A-configuration in 
spectral-line mode. This data has been published by its original 
investigators (Werner et al. 2001; Werner 2002) and 
was processed similarly. In brief, we used AIPS (Bridle \& Greisen 
1994) for the initial calibration utilizing scans of 3C~286 to set the 
flux densities to the \vla\ 1999.2 scale. The nearby source 1800+784 
was used as the phase calibrator for \ngc, and bandpass calibration 
performed using 3C~286. The inner 22/31 channels were averaged to 
produce a single $\sim$9 MHz bandwidth continuum dataset centered at 
about 1.4 GHz which was then self-calibrated in the Caltech DIFMAP 
package (Shepherd et al. 1994). The final images were 
corrected for primary beam response using the AIPS task PBCOR. This 
correction amounts to up to an $\sim$5\% correction near the 
$\sim$4\arcmin\ distant X-ray feature in the jet.

\section{Results}   
\begin{figure}  
\psfig{figure=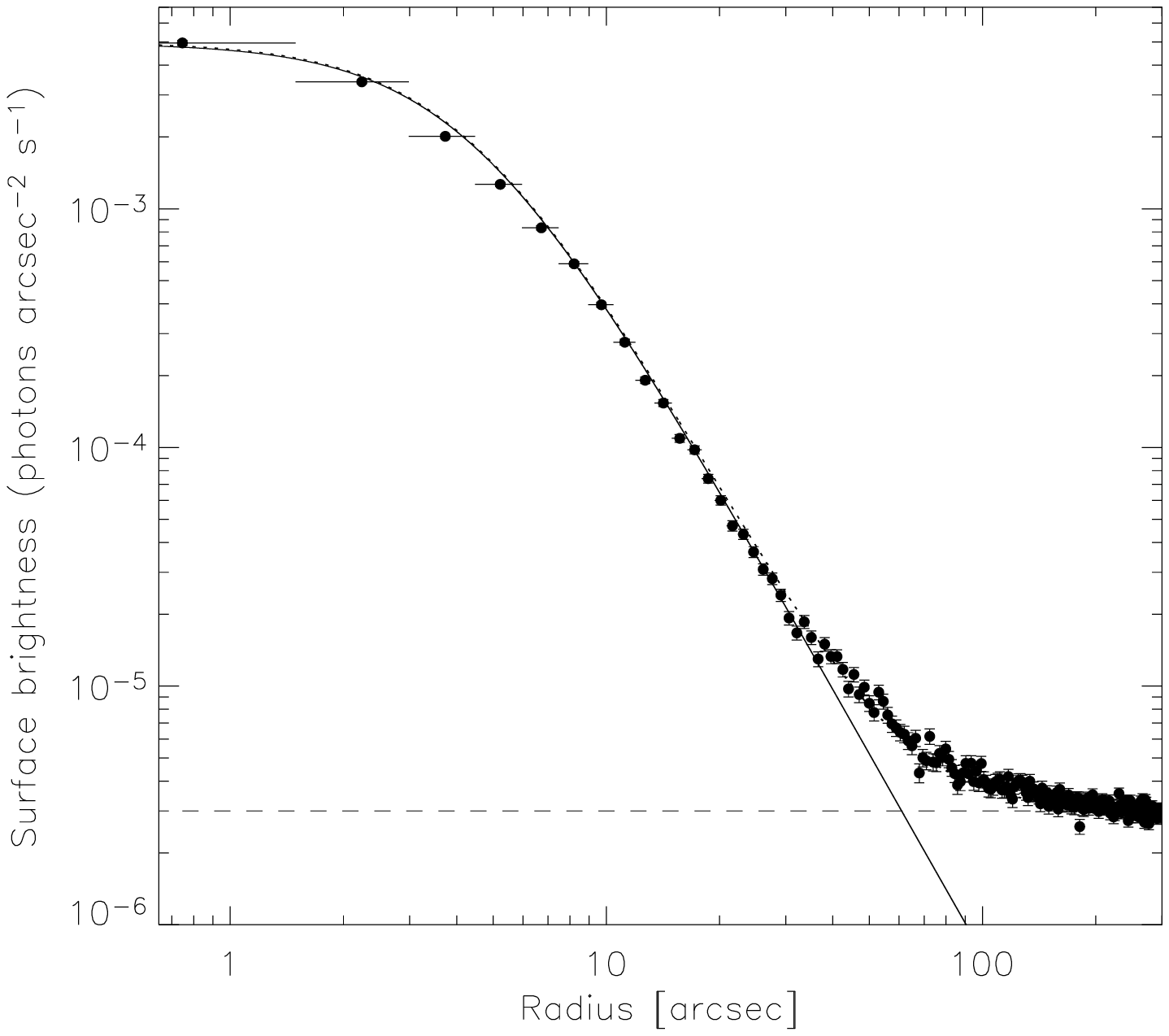,height=6cm,width=8.5cm,%
bbllx=0pt,bblly=60pt,bburx=535pt,bbury=435pt,angle=0,clip=}
\psfig{figure=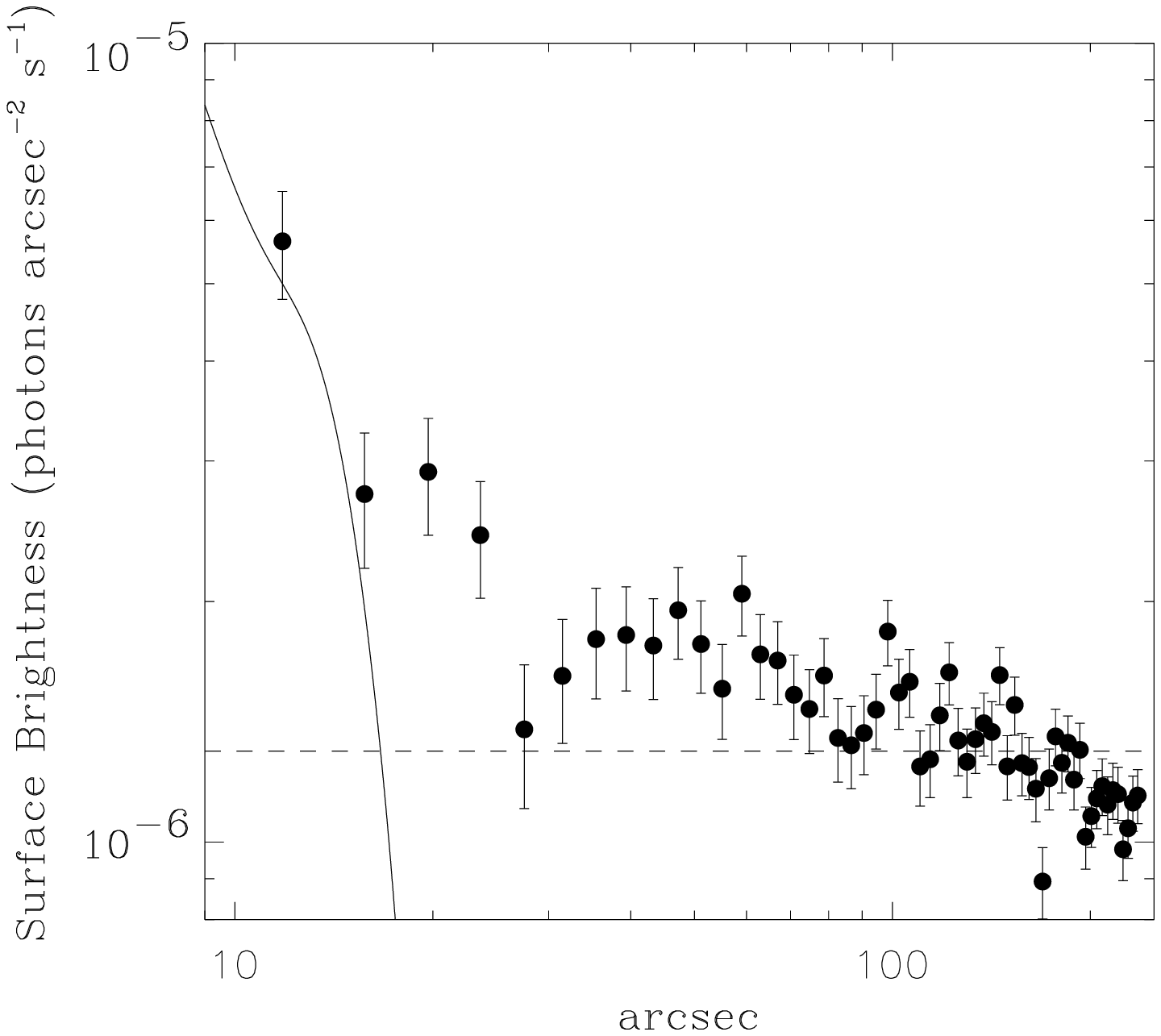,height=6cm,width=8.5cm,%
bbllx=0pt,bblly=60pt,bburx=535pt,bbury=435pt,angle=0,clip=}
\psfig{figure=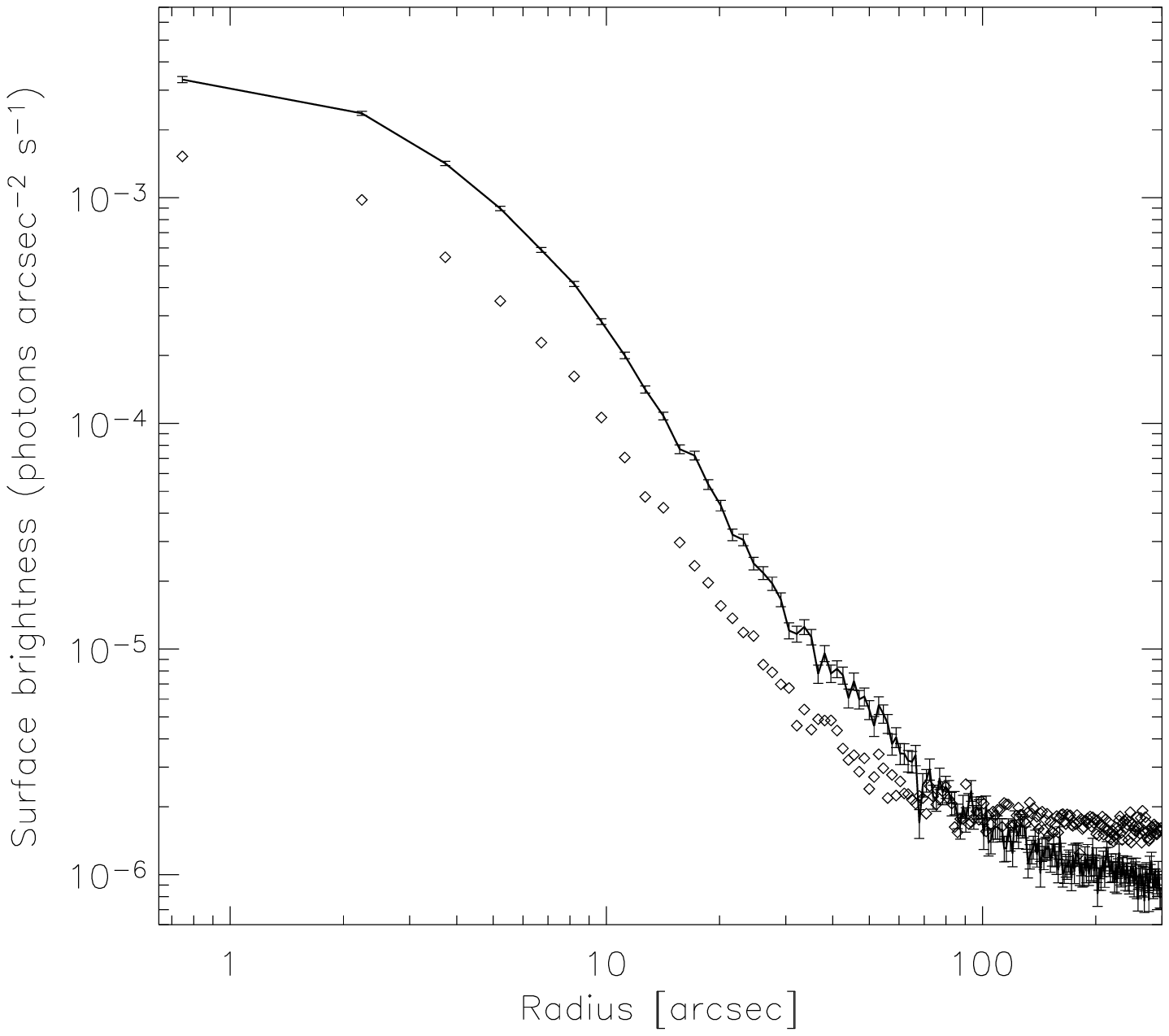,height=6cm,width=8.5cm,%
bbllx=0pt,bblly=60pt,bburx=535pt,bbury=435pt,angle=0,clip=}
\caption{Radial profiles of \ngc. {\it (a, Top):} EPIC MOS2 profile in 
total energy band 0.3--10 keV compared to the instrumental PSF (solid
line; see text) and background (dashed line). {\it (b, Middle):}
ACIS-I radial profile, compared to the instrumental PSF (solid
line). A clear excess over the PSF is present between
20--100\arcsec. {\it (c, Bottom):}
Comparison of the EPIC MOS2 profile at soft
(0.3--2 keV, solid line plus errors) and hard (2--10 keV, open
diamonds) X-rays.  The soft and hard X-ray profiles are similar.}
\end{figure}   
The EPIC pn image of \ngc\ in 0.3--10 keV is shown in Fig. 1a, with 
the \vla\ contours at 1.4 GHz superposed. The image was binned by a 
factor 32 and smoothed with a Gaussian function with width 
$\sigma$=6\arcsec. The final resolution is $\sim$10\arcsec. 
The smoothed ACIS-I image in 0.3--8 keV is shown in Fig. 
1b. This image was binned by a factor 4 and smoothed with the 
\verb+CSMOOTH+ with final resolution 2\arcsec. The radio images 
overlaid in the figures were restored with beamsizes matching the 
resolutions of the respective X-ray images. 
  
As apparent in Fig. 1a-b, various components contribute to the total 
X-ray emission of the galaxy: the nucleus, a diffuse isotropic halo 
associated with the elliptical galaxy (e.g., Birkinshaw \& Worrall 
1993), and jet emission at $\sim$ 4\arcmin\ from the core.  Also 
apparent in the \chandra\ image are several point sources scattered as 
far as 4--5\arcmin\ from the nucleus (see Appendix). Using the 
\chandra\ data, we determined that their contribution to the total 
X-ray emission in the EPIC extraction radius of the halo is \ltsima 
5\%. 
  
\subsection{The X-ray halo}  
\begin{figure}
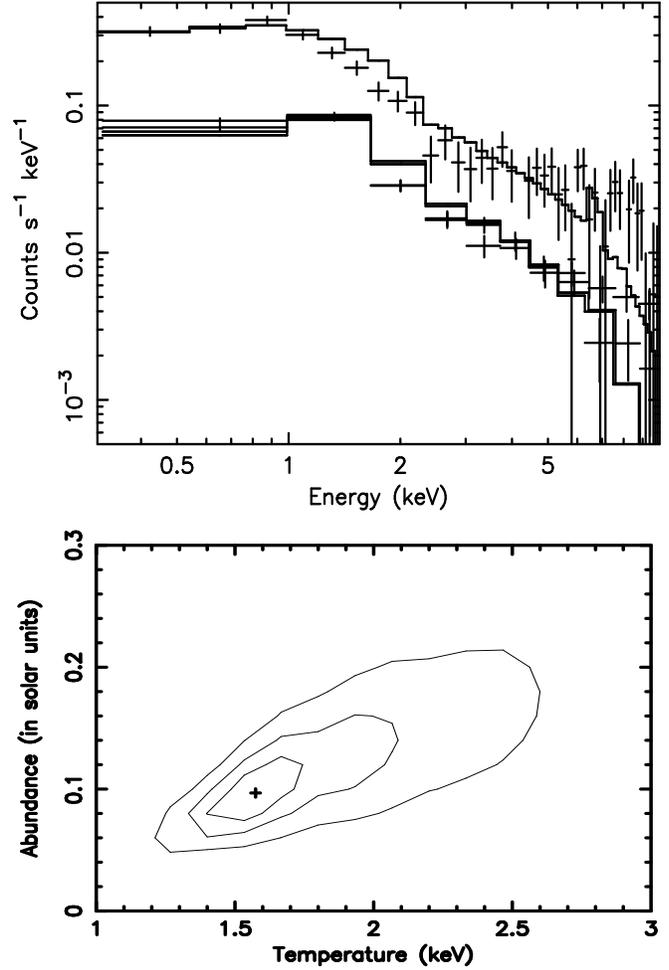
  
\psfig{figure=h4680f3a.ps,height=7.cm,width=8.7cm,%
bbllx=132pt,bblly=98pt,bburx=535pt,bbury=628pt,angle=-90,clip=}  
\psfig{figure=h4680f3b.ps,height=6.0cm,width=8.7cm,%
bbllx=118pt,bblly=98pt,bburx=535pt,bbury=635pt,angle=-90,clip=}   
\caption{{\it Top:} EPIC pn+MOS spectra of the halo of \ngc, extracted in an  
annulus of radii 35-215\arcsec. In the top panel, the top spectrum is 
the pn, the bottom two datasets are the MOS1 and MOS2 data. The data 
were fitted with a thermal model; excess flux is apparent above 3 keV, 
suggesting a harder component. {\it Bottom:} Confidence contours at 
68\%, 90\%, and 99\% confidence for the temperature and abundance of 
the thermal component, from the fits to the EPIC pn+MOS data.} 
\end{figure}   
To investigate the spatial properties of the halo, we extracted
brightness radial profiles from EPIC MOS2 on-board \xmm\ and ACIS-I
on-board \chandra. The choice of the MOS2 detector is motivated
by geometrical considerations, namely, the fact that the source is
centered on its CCD1, and its higher sensitivity than MOS1, offering
the best chance to study the extended emission. The EPIC radial
profile in the total energy band 0.3--10 keV is shown in Fig. 2a,
together with the instrumental PSF and the background, while the
ACIS-I profile is shown in Fig. 2b. The MOS2 profile is fully
consistent with the instrumental PSF (the MOS2 PSF was obtained using
the analytical formula from Ghizzardi 2001) plus background, whereas
the ACIS profile suggests the presence of an excess between 20\arcsec\
and 100\arcsec. However, the limited signal-to-noise ratio of the ACIS
data hampers a more quantitative spatial analysis. We fitted the
\chandra\ profile with a $\beta$ model; the fitted parameters are  
unconstrained.   
 
Fig. 2c shows the energy-dependent EPIC profiles at soft 
(0.3--2 keV) and hard (2--10 keV) X-rays. The soft and hard X-ray 
profiles are similar, suggesting that the the hard X-ray emission 
originates within the galaxy.  
  
EPIC pn and MOS spectra for the halo were extracted as described 
above. We first fitted the total integrated spectrum with a thermal 
plasma model (\verb+apec+ in \verb+XSPEC+), leaving as free parameters 
the temperature, $kT$, the abundance, $Z$ (in solar units), and the 
normalization. All components are absorbed by a column density fixed 
to the Galactic value, $N_{\rm H}^{\rm Gal}=5.5 \times 10^{20}$ 
\nh\ (Dickey \& Lockman 1990). The fit gives $\chi_{\rm r}^2=1.2/2913$.   
Inspection of the residuals shows the presence of excess flux above 
the model at energies \gtsima 3 keV (Fig. 3a). Adding a power law to 
the model gives $\chi^2_{\rm r}$=1.10/2911, improved at $P_{\rm F} >$ 99\% 
with 
respect to the fit with the thermal alone. When the hard tail is 
modeled with a thermal bremsstrahlung component, the fit is worse 
($\chi^2_{\rm r}=1.14/2911$) than with a power law.  The best-fit 
temperature is 199 keV, effectively mimicking a flat power law. 
  
The fitted parameters of the thermal plus power law model are: 
$kT=1.57 \pm 0.09$ keV, $Z=0.10 \pm 0.03 Z_{\odot}$, and photon index 
$\Gamma=-0.08^{+0.08}_{-0.3}$. The confidence contours for the 
temperature and abundance of the thermal component are shown in Fig. 
3b. The observed 0.5--2 keV flux of the thermal is F$_{\rm 0.5-2~keV} 
\sim 3.1-4.7 \times 10^{-12}$ \flux, and the intrinsic 
(absorption-corrected) 0.5--2 keV luminosity is L$_{\rm 0.5-2~keV} \sim 
4-6 \times 10^{42}$ \lum. The power law accounts for $\sim$ 20\% of 
the total X-ray emission in 0.5--2 keV. Its observed flux and 
intrinsic luminosity in 2--10 keV are F$_{\rm 2-10~keV} 
\sim 1.7-2.6 \times 10^{-11}$ \flux, and L$_{\rm 2-10~keV} \sim 1.8-2.7 
\times  10^{42}$ \lum, respectively.  
  
We investigated whether the hard power law is related to the 
contaminating wings of the PSF (the nucleus is a bright X-ray emitter; 
Paper I). We extracted the halo spectra in annuli of increasing radii, 
as described in Sect. 2.1. If these spectra are fitted with a thermal 
model only, the fit is acceptable ($\chi^2_{\rm r}=0.9-1.2$) but the fitted 
temperature is very high, $kT=4-5$ keV, much larger than typically 
found for the halo of radio galaxies, $kT \sim 1$ keV (e.g., Worrall, 
et al. 2001). Inspection of the residuals shows 
that a power law is present even in the spectrum extracted in the 
outermost annulus between 155\arcsec\ and 215\arcsec. If the power law 
is added, the fits improve with high significance, $P_{\rm F}$ \gtsima 99\%; 
its photon index is not well defined and ranges between --0.7 and 
1.2. Within the uncertainties, there is no variation of either the 
temperature or the abundance of the thermal component with distance 
from core. The temperature varies in the range 1.3--1.9 keV and the 
abundance in 0.05--0.2 (90\% uncertainties).  Moreover, the nucleus 
spectrum has a substantially steeper slope, $\Gamma \sim 1.9$ (Paper 
I), than the hard component. It is thus unlikely that the power law 
represents the contribution of the PSF wings. 
  
A non-thermal component is also required to fit the \chandra\ spectrum 
of the halo. The latter was extracted in a similar region as for EPIC, 
an annulus of radii 35\arcsec\ and 215\arcsec, after removing the jet 
region and the point sources. Spectra were extracted for each CCD and 
fitted jointly. A fit with a thermal only is good 
($\chi^2_{\rm r}=0.96$/209), but yields an unacceptably high temperature $kT 
\sim 10$ keV for $Z=0.08$. Adding a power law causes the $\chi^2$ to 
decrease by $\Delta\chi^2$=35, significant at $>$ 99.9\% 
confidence. The fitted temperature then converges towards the EPIC 
value with $kT=1.25^{+0.36}_{-0.21}$ keV. 
  
To investigate the spatial distribution of the hard X-ray component,  
we extracted soft (0.5--2 keV) and hard (2--8 keV) ACIS  
images. The hard X-ray image is dominated by the nucleus point source  
and there are not enough counts in the diffuse emission to study its  
spatial properties.   
  
In summary, the EPIC spectrum of the diffuse halo in \ngc\ is 
best-fitted by a thermal model with temperature $\sim$ 1.6 keV and 
sub-solar abundances. There are no gradients of either parameter with 
distance from the core. The data also require a hard X-ray component 
above 3 keV, which is best-fitted by a power law with a flat slope, 
albeit poorly determined. The hard component, which accounts for 20\% 
of the total 0.5--10 keV luminosity of the halo, is also detected in 
the \chandra\ spectrum.  
  
\subsection{X-ray emission from the jet}   
  
Fig. 1a-b shows that weak X-ray emission is detected at $\sim$ 
4\arcmin\ from the core in both the EPIC and ACIS images, coincident 
with the bright radio jet. To assess the significance of the jet 
detection we compared the average counts found in the elliptical jet 
region with the counts found in six identical elliptical regions 
randomly located around the jet. The significance of the jet detection 
is of the order of 10$\sigma$.  We thus confirm the previous claim 
based on \rosat\ of X-ray emission from the outer jet (Mack et 
al. 1997a). The X-ray emission of the jet is diffuse in both the EPIC 
and ACIS images, with no compact knots at the resolution of ACIS-I. 
  
We extracted X-ray spectra of the jet with EPIC. The pn and MOS 
spectra were fitted with a single power law model with Galactic 
absorption, $5.5 \times 10^{20}$ \nh. This fit is acceptable, 
$\chi_{\rm r}^2$=0.98 for 102 degrees of freedom, and gives a best-fit 
photon index $\Gamma=1.15^{+0.38}_{-0.35}$. An equally acceptable fit 
was obtained with a thermal model; however, the fitted temperature is 
$kT=63$ keV and the abundances near zero, effectively mimicking a 
power law. A fit with a thermal plus a power-law model does not lead 
to significant improvement. Using the best-fit power law model, the 
observed 2--10 keV flux is $F_{\rm 2-10~keV}=(1.6-3.1) \times 10^{-14}$ 
\flux, and the 1 keV flux is $F_{\rm 1~keV}=2.3^{+0.7}_{-0.6}$ nJy. 
  
A similar result is obtained with \chandra. The ACIS-I spectrum of the 
jet was extracted in a 38\arcsec $\times$ 22\arcsec\ elliptical 
region. In the energy range 0.3--8 keV, about 109 counts were derived, 
allowing a crude spectral analysis. Again, the model that best-fits 
the ACIS data is a power law, with $\Gamma=1.06^{+0.42}_{-0.38}$ and 1 
keV flux F$_{\rm 1~keV}=3.1^{+2.5}_{-1.0}$ nJy. These parameters are 
consistent with \xmm\ within the uncertainties. As for the EPIC 
spectra, a thermal model is again formally acceptable but the fit 
converges to a temperature $kT \sim 68$ keV and near-zero abundances. 
  
In summary, X-ray emission from the jet at 4\arcmin\ from the core  
is detected with both \xmm\ and \chandra. The X-ray spectrum of the  
jet is better fitted by a flat ($\Gamma \sim 1.2$) power law.   
  
\section{Discussion}   
  
\subsection{The X-ray halo}   
  
\ngc\ is embedded in a diffuse X-ray halo extending \gtsima  
100\arcsec, or \gtsima 45 kpc from the core.  The fitted temperature  
of the halo, $kT \sim 1.6$ keV, is significantly larger than derived  
from \rosat\ PSPC data, $\sim$ 0.5 keV (Birkinshaw \& Worrall  
1993). This is not surprising, as \ngc\ is thought to host a cooling  
flow in the nuclear region (Birkinshaw \& Worrall 1993), which was  
excluded from our extraction region. Indeed, spectral fits to the EPIC  
spectrum of the nucleus requires a thermal component with a  
temperature $kT \sim 0.55$ (Paper I), in agreement with the \rosat\  
data.   
  
A new result of our analysis is the detection of a hard X-ray 
component in the spectrum of the halo above 3 keV. The hard tail is 
best described by a power law, contributing $\sim$ 20\% of the total 
0.5--10 keV X-ray flux. A thermal model for the X-ray tail is 
inconsistent with the data. The hard non-thermal component is 
independently confirmed by an archival \chandra\ observation.  
  
Inspection of the \chandra\ image (Fig. 1b) shows that the hard 
component in the EPIC spectrum cannot be due to the collective 
contribution of the detected off-axis X-ray point sources. The hard 
X-ray component is still present in the ACIS data when the point 
sources are excised. The hard X-ray profile follows closely the soft 
profile (Fig. 2c), indicating that the hard X-ray emission is related 
to the galaxy halo.  
 
We thus investigated the possibility that the hard X-ray component 
could be due to the emission of a population of undetected Low-Mass 
X-ray Binaries (LMXBs). Based on recent \chandra\ observations of 
nearby ellipticals (e.g., NGC~1316, Kim \& Fabbiano 2003; NGC~5128, 
Kraft et al. 2003; NGC~1399, Angelini et al. 
2001), this contribution could be substantial. First, we examined the 
nature of the detected off-axis point sources in Table 1. Using the 
count rates reported in the Table, and assuming a power law model with 
$\Gamma=1.5-2.0$ and Galactic absorption, we estimate that the 
threshold X-ray flux in 0.3--8 keV for detecting point sources is $3 
\times 10^{-15}$ \flux. This corresponds to a luminosity 
L$_{\rm 0.3-8~keV} \sim 3 \times 10^{39}$ \lum, at least one order of 
magnitude larger than for LMXBs.  
 
Based on the luminosity functions of field sources from X-ray deep 
surveys (Kim et al. 2003), in a 4\arcmin\ radius centered on \ngc\ we 
expect 14 $\pm$ 4 serendipitous sources at soft X-rays, consistent 
with our measured rate of 10 sources (Table 1). It is thus likely that 
the off-axis point sources in Table 1 are field galaxies, although we 
cannot exclude the possibility that some are Ultra-Luminous X-ray 
sources, which indeed were found in other ellipticals (e.g., Jeltema 
et al. 2003).  
 
If a population of LMXBs is indeed present in \ngc, it is undetected
in our \xmm\ and \chandra\ images. However, its presence could affect
the integrated spectrum. To check for this, we added a new component
to the best-fit model in Sect. 3.1 mimicking the integrated contribution
of the LMXBs. Following Kim \& Fabbiano (2003), we modeled this
component with either a steep ($\Gamma \sim 1.8$) power law, or a $kT
\sim 5$ keV thermal model. The addition of either the power law or
thermal does not improve the fit to the halo spectrum. Moreover, the
spectral parameters of the best-fit model components (Sect. 3.1) are
basically unaffected.
 
An independent indicator of the relative contribution of LXMBs to the 
X-ray emission of the galaxy is given by the ratio of the 
X-ray-to-optical luminosities, $L_{\rm X}/L_{\rm B}$, where $L_{\rm B}$ is 
in solar units. Kim et al. (1992) showed that the LMXB 
contribution decreases with increasing  $L_{\rm X}/L_{\rm B}$. For \ngc, we 
derive $\log(L_{\rm X}/L_{\rm B})\simeq$32.3, which according to Kim et 
al. (1992) 
indicates that the LMXB contribution to the hard X-ray emission is 
negligible. This is confirmed by the later analysis of Matsumoto et 
al. (1997), who analyzed \asca\ observations of 12 early-type 
galaxies. The X-ray halo luminosity of \ngc\ is two orders of 
magnitude larger than predicted by the $L_{\rm X}-L_{\rm B}$ relationship in 
Fig. 6 of Matsumoto et al. (1997). We assumed the halo luminosity 
given in Sect. 3.1, and an optical magnitude $B=13.64$ mag (de 
Vaucouleurs et al. 1991). Based on this evidence, we conclude that it 
is unlikely that the hard X-ray tail in the EPIC spectrum of the \ngc\ 
halo is due mostly to unresolved LMXBs. 
 
The low abundances we detect for the halo are puzzling. While in some 
sources (e.g., NGC~1316; Kim \& Fabbiano 2003) they seem to be related 
to LMXB contamination of the integrated diffuse emission, this is not 
the case for \ngc. Deeper X-ray observations designed to probe the 
structure of the diffuse thermal emission are needed.   
  
We explored alternative origins for the hard X-ray component. There
are theoretical reasons to believe that non-thermal halos should be
present in clusters/groups containing radio galaxies, as the latter
are important sources of energetic particles (e.g., Ensslin et
al. 1997). The energy stored in intergalactic magnetic fields can be
large and will last long after the activity in the nucleus is turned
off (Kronberg et al. 2001), heating the ICM through dissipation
processes. Interestingly, in \ngc\ the \gtsima 35\arcsec\ halo
temperature, $kT \sim$ 1.6 keV, is hotter than in other ``normal''
giant elliptical galaxies (e.g., Xu et al. 2002) and to other less
powerful radio galaxies (e.g., NGC~4261; Gliozzi et al. 2003b), where
$kT \sim 0.5-0.7$ keV.  This lends credit to the idea that the AGN can
inject particles and energy heating the ISM.
  
The hard component detected in \ngc\ is similar to the one previously 
observed with \asca\ in the nearby group of galaxies HCG~62 (Fukazawa 
et al. 2001). The latter authors find that either a power law with 
photon index $\Gamma=0.8-2.7$, or a very hot ($kT > 6$ keV) 
bremsstrahlung best-fits the data at energies $>$ 4 keV, in both cases 
accounting for $\sim$ 20\% of the total 0.5--10 keV flux of the 
halo. Fukazawa et al. (2001) conclude that the most likely origin of 
the hard X-ray component is either inverse Compton scattering of the 
Cosmic Microwave Background photons (IC/CMB) on a population of 
low-energy ($\gamma \sim 10^3$) electrons, if the magnetic field of 
the ICM is less than $1\mu$Gauss, or non-thermal bremsstrahlung from a 
population of energetically important relativistic electrons. More 
recently, non-thermal diffuse emission was detected in a deep 
\chandra\ image of the distant ($z$=3.8) radio galaxy 4C 41.17 (Scharf 
et al. 2003), and interpreted as IC on the local CMB and FIR radiation 
fields. 
\begin{figure}  
\centerline{\psfig{figure=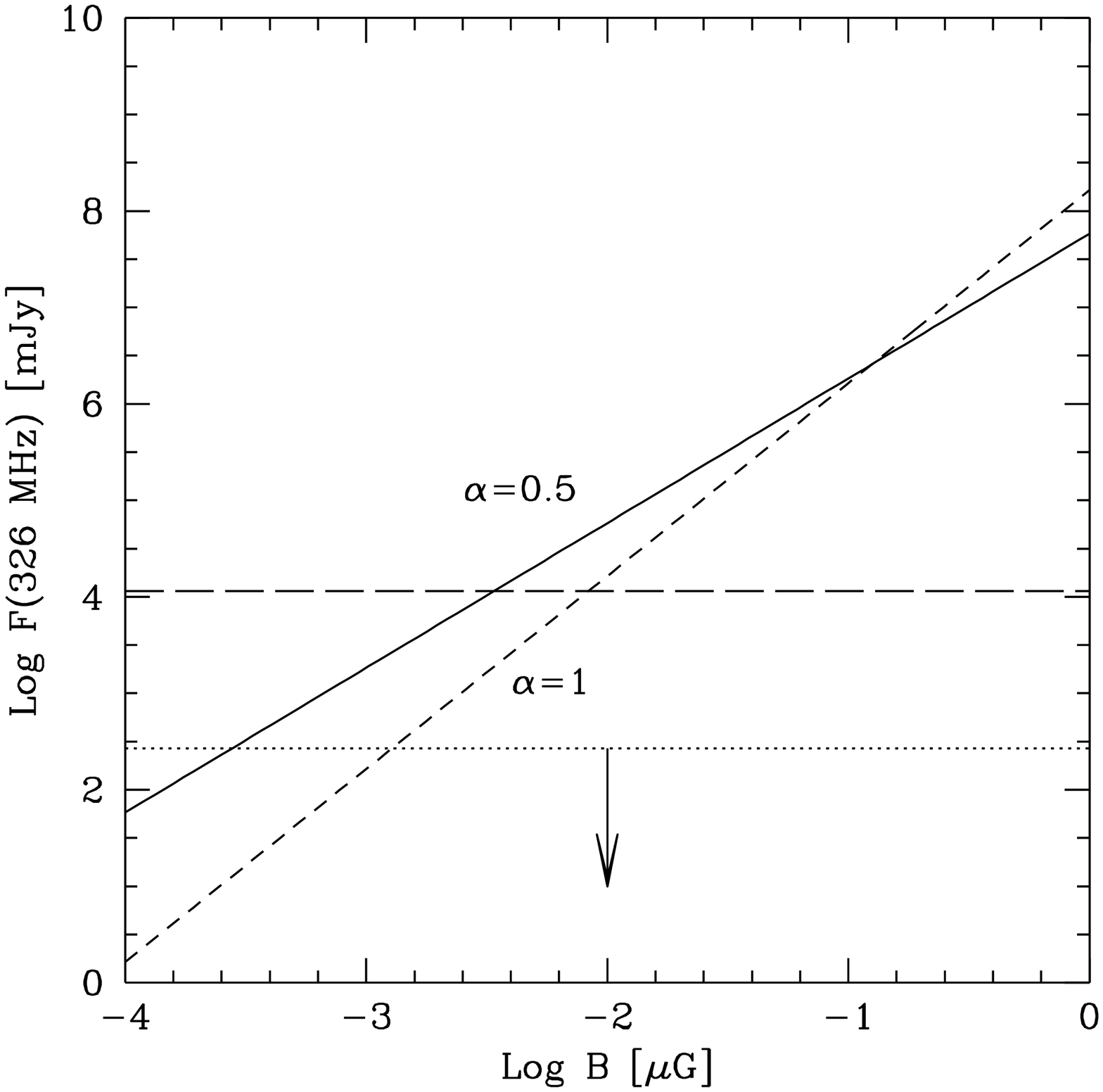,height=7.2cm,width=8.7cm}}  
\caption{Plot of the predicted synchrotron flux at 326 MHz versus  
magnetic field for the halo, assuming two different slopes for the 
radio spectrum. The horizontal dotted and dashed lines are the upper 
limits to the halo and total fluxes at 326 MHz from the WSRT map of 
Mack et al. (1997b).} 
\end{figure}    
Following Fukazawa et al. (2001), a possible origin for the X-ray tail 
in \ngc\ is IC/CMB off a population of old electrons. This possibility 
is particularly attractive considering that \ngc\ has an active radio 
core and a long jet which disrupts at $\sim$ 4\arcmin\ from the core, 
where diffuse X-ray emission is detected. If the jet magnetic field is 
disrupted and becomes unable to confine the electrons, the latter 
could have escaped and diffused in the ICM of the galaxy. The 
relativistic electrons would then lose energy via IC scattering of the 
CMB photons, whose energy density at the redshift of \ngc\ is 
non-negligible (see below). 
  
However, the very flat slope of the hard X-ray component we measure 
from both \xmm\ and \chandra\ spectra, $\Gamma \sim 0$ (albeit with 
large uncertainties), posits a problem for the IC/CMB 
interpretation. In fact, this implies that the {\it minimum} Lorentz 
factor of the scattering electrons is large, $\gamma_{\rm min} \sim 
3000$. We would thus be observing the low-energy Compton tail emission 
(characterized by a spectral slope $\alpha =-1$) from a very energetic 
electron population. 
  
Kronberg et al. estimate for \ngc\ a minimum total energy content of 
the X-ray halo of $\sim 4.2 \times 10^{59}$ ergs and a {\it minimum} 
halo magnetic field of $\sim 4\mu$Gauss. An immediate consequence of 
such large magnetic field and energetic electrons is that one should 
observe large radio fluxes from the halo.  
 
A deep 326 MHz Westerbork Synthesis Radio Telescope (WSRT) map (Mack
et al. 1997b) shows no obvious sign of such a radio halo surrounding
this galaxy. The only diffuse emission in the image is detected in the
actual radio lobes near the edges of the source. We estimate an
integrated flux limit of 275 mJy (3 sigma) for the halo out to
215\arcsec\ (the extent of the \xmm\ extraction aperture) from
measuring the RMS in this region of the digital FITS version 326 MHz
map of the Mack et al. (1997b) which was available to download from
DRAGN web-site\footnote{J.~P. Leahy, A.~H.  Bridle, \& R.~G. Strom,
http://www.jb.man.ac.uk/atlas/}.  Fig. 4 shows the plot of the
predicted synchrotron flux at 326 MHz versus the halo magnetic field
for two different values of the radio spectral index. In this
calculation we assume that the observed hard X-ray power-law is due to
the low-energy tail of electrons at $\gamma _{\rm min} \sim 3000$. Above
an energy $E\sim 10$ keV the (unobserved) emission would be a
power-law with slope $\alpha$, produced by electrons with $\gamma
>\gamma_{\rm min}$. In order to reproduce the observed radio flux
(horizontal dotted line), a magnetic field $B_{\rm halo}$ \ltsima 10$^{-3}
\mu$Gauss is required.  This is a factor 10 or more smaller than
generally measured for radio galaxies (Feretti et al. 1995). However,
measurements of rich clusters with \sax\ and \rxte\ suggest that
intergalactic magnetic fields can be very weak (e.g., Valinia et
al. 1999). Note that, however, for magnetic fields below $B \sim 1
\mu$G, if electrons have a high energy cut-off at $\gamma_{\rm max} <
10^4$ the maximum frequency of the synchrotron emission will be
located below $\sim 300 B_{\rm 1 \mu G} \gamma^2_{\rm max, 4}$ MHz. Therefore
if the high-energy non-thermal electron population has a narrow
electron distribution, $3000 < \gamma < 10^4$, the synchrotron
emission would be unobservable at $\sim $ 300 MHz even with moderately
low magnetic fields, $B$ \ltsima $1\mu$G.  A similar solution
(IC/CMB scattering off low-energy electrons, emitting synchrotron
radiation well below the observed frequencies) was independently
proposed for the high-redshift ($z=4.3$) quasar GB1508+5714, where an
X-ray halo with no radio counterpart was recently detected (Yuan et
al. 2003).

Alternatively, as discussed by Fukazawa et al. (2001) for HCG~62, the  
X-ray tail could be due to non-thermal bremsstrahlung involving  
subrelativistic but suprathermal electrons (e.g., Kempner \& Sarazin  
2000). Assuming the ratio of thermal to non-thermal luminosity of 0.2,  
Fukazawa et al. derive that the energy density of non-thermal  
electrons would be $\sim$ 0.6 times that of thermal electrons. Similar  
conclusions apply in the case of  
\ngc, where the non-thermal component accounts for a similar fraction  
of the diffuse X-ray flux. Kempner \& Sarazin (2000) discuss a model  
where the non-thermal X-ray tail in the cluster Abell 2199 is produced  
by a population of electrons distributed in energy as a power law of  
index $\mu$. For a steep distribution ($\mu$ \gtsima 3.5), the X-ray  
tail can be approximated by a power law with $\Gamma \sim 1 +  
0.5\mu$. In our case, we would derive $\mu \sim -2$, implying the  
presence of a large number of very energetic electrons. As these  
electrons would lose energy via synchrotron, again one would expect to  
observe diffuse radio emission.   
  
In summary, we find evidence for the presence of a non-thermal  
high-energy tail in the halo of \ngc\ with both \xmm\ and  
\chandra. The physical origin of this component, however, is not  
clear. A possible explanation - IC/CMB off electrons in the halo -  
implies very weak magnetic fields.  Non-thermal bremmstrahlung 
requires the presence of a large number of very energetic electrons.  
  
\subsection{The X-ray jet}  
  
The EPIC and ACIS data confirm the previous findings from \rosat\ that 
the outer jet emits X-rays. The X-ray spectrum of the jet is best 
described by a power law with $\Gamma \sim 1.2$, while a thermal model 
is formally ruled out by the data. Fig. 5 shows the radio-to-X-ray 
spectral energy distribution of the jet. The radio flux at 1.4 GHz was 
measured (see above) using the archival \vla\ image and an extraction 
region similar to the EPIC extraction region. The radio spectrum 
between 1.4 and 5 GHz throughout the jet was determined to be 
$\alpha_{\rm r}=0.64\pm0.05$ by Perley et al. (1984). We adopted this 
spectrum and used our 1.4 GHz measurement to calculate representative 
flux values at 1.66 and 4.885 GHz for the region of interest. The open 
bowtie represents the best-fit EPIC spectrum. Extrapolating the radio 
spectrum to the X-ray region (dotted line in Fig. 5) overestimates the 
X-ray flux by more than 2 orders of magnitude. Moreover, the X-ray 
spectrum is much flatter than the lower-energy extrapolation. This 
suggests that the X-ray emission is not the high-energy tail of the 
synchrotron responsible for the longer wavelengths, and that a 
different spectral component is needed to account for the X-rays. 
\begin{figure}  
\centerline{\psfig{figure=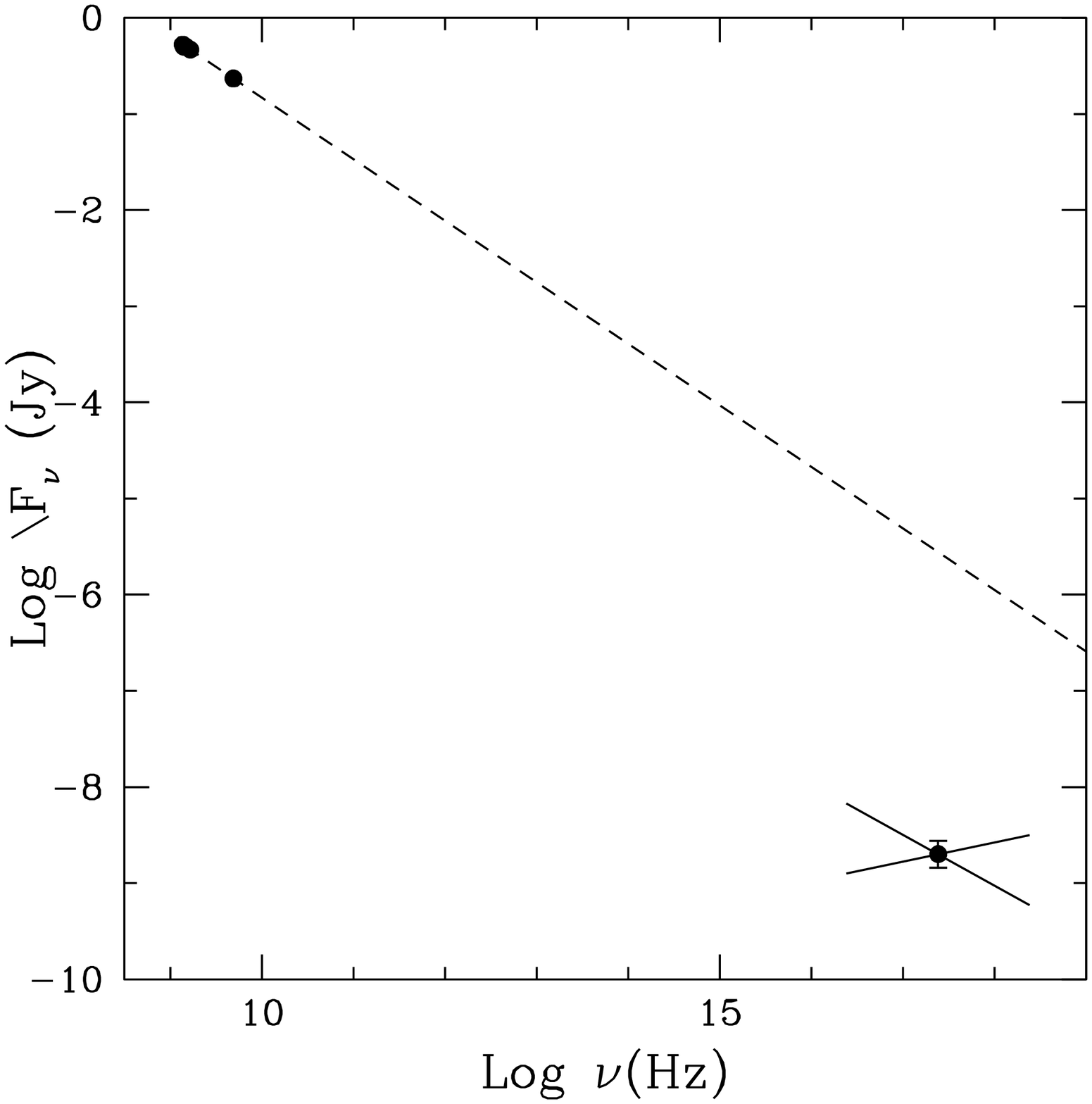,height=7.2cm,width=8.7cm}}  
\caption{Spectral energy distribution of the outer jet. The dotted  
line is the extrapolation of the radio spectrum, which overestimates  
the measured X-ray flux by more than two orders of magnitude.} 
\end{figure}   
    
Based on the \rosat\ PSPC detection, Mack et al. (1997a) argue in
favor of a thermal origin for the X-ray emission. However, a thermal
origin for the X-ray emission is ruled out independently by the EPIC
and ACIS spectra (see Sect. 3.2). We also note that there is no evidence
for internal depolarization from the radio (Perley et al. 1984). This
makes a thermal origin for the X-rays unlikely.  However, vis-a-vis
the limited signal-to-noise ratio of the EPIC/ACIS spectra, the
possibility that at least a fraction of the X-rays are due to thermal
emission from ambient gas remains open.
  
A likely possibility is inverse Compton (IC) scattering. Seed photons
for IC can be provided by 1) the synchrotron photons themselves (SSC
process); 2) the beamed emission from the inner jet; 3) the isotropic
emission of the AGN; and 4) the Cosmic Microwave Background radiation
(CMB). At a projected distance of 94 kpc from the nucleus, the Narrow
Line Region and dusty torus are not a significant source of photons.
  
We compared the radiation densities (in the blob comoving frame) at 
the blob location for processes 1--4 in order to evaluate the dominant 
source of photons. The SSC luminosity was obtained integrating the 
radio spectrum in Fig. 5 up to $10^{12}$ Hz, $L_{\rm s} \sim 1.9 \times 
10^{41}$ \lum. Assuming a size for the X-ray emission region of 
10\arcsec\ (the EPIC resolution), the synchrotron radiation density is 
$u_{\rm s} \sim 2.8 \times 10^{-15}\delta^{-4}$ erg cm$^{-3}$. Here $\delta$ 
is the jet Doppler factor, defined as 
$\delta=[\Gamma_{\rm L} (1-\beta\cos\theta)]^{-1}$, with $\Gamma_{\rm L}$ the 
jet Lorentz factor, $\beta$ the plasma speed, and $\theta$ the jet 
inclination to the line of sight.  
 
The contribution from the inner jet (e.g., Celotti et al. 2001) can be 
evaluated as $u_{\rm in}\simeq 3\times 10^{-15} L_{\rm in,46}, \Gamma 
^{-2}_{\rm in,1} \Gamma_{\rm L}^{-2}$ erg cm$^{-3}$, where $L_{\rm in}$ 
is the typical 
observed luminosity of blazars (e.g., Ghisellini et al. 1998) and 
$\Gamma_{\rm in}$ is the Lorentz factor of the inner jet. 
  
For the AGN luminosity, we used the bolometric luminosity from Paper  
I, $L_{\rm AGN} \sim 4 \times 10^{43}$ \lum. At the location of the X-ray  
jet, the radiation density due to the isotropic radiation of the AGN  
is $u_{\rm AGN} \sim 1.3 \times 10^{-15}\Gamma_{\rm L}^{-2}$ erg cm$^{-3}$.  
  
The CMB radiation density scales like $(1+z)^4$ and $\Gamma_{\rm L}^2$. We 
find $u_{\rm CMB} \sim 4.4 \times 10^{-13} \Gamma_{\rm L}^2$ erg 
cm$^{-3}$. This is at least 2 orders of magnitude larger than $u_{\rm s}$, 
$u_{\rm in}$, and $u_{\rm AGN}$. Thus, the CMB is the dominant source of seed 
photons for the IC process. Note that assuming significant beaming 
increases the importance of $u_{\rm CMB}$ and decreases that of the other 
processes. 
\begin{figure}  
\centerline{\psfig{figure=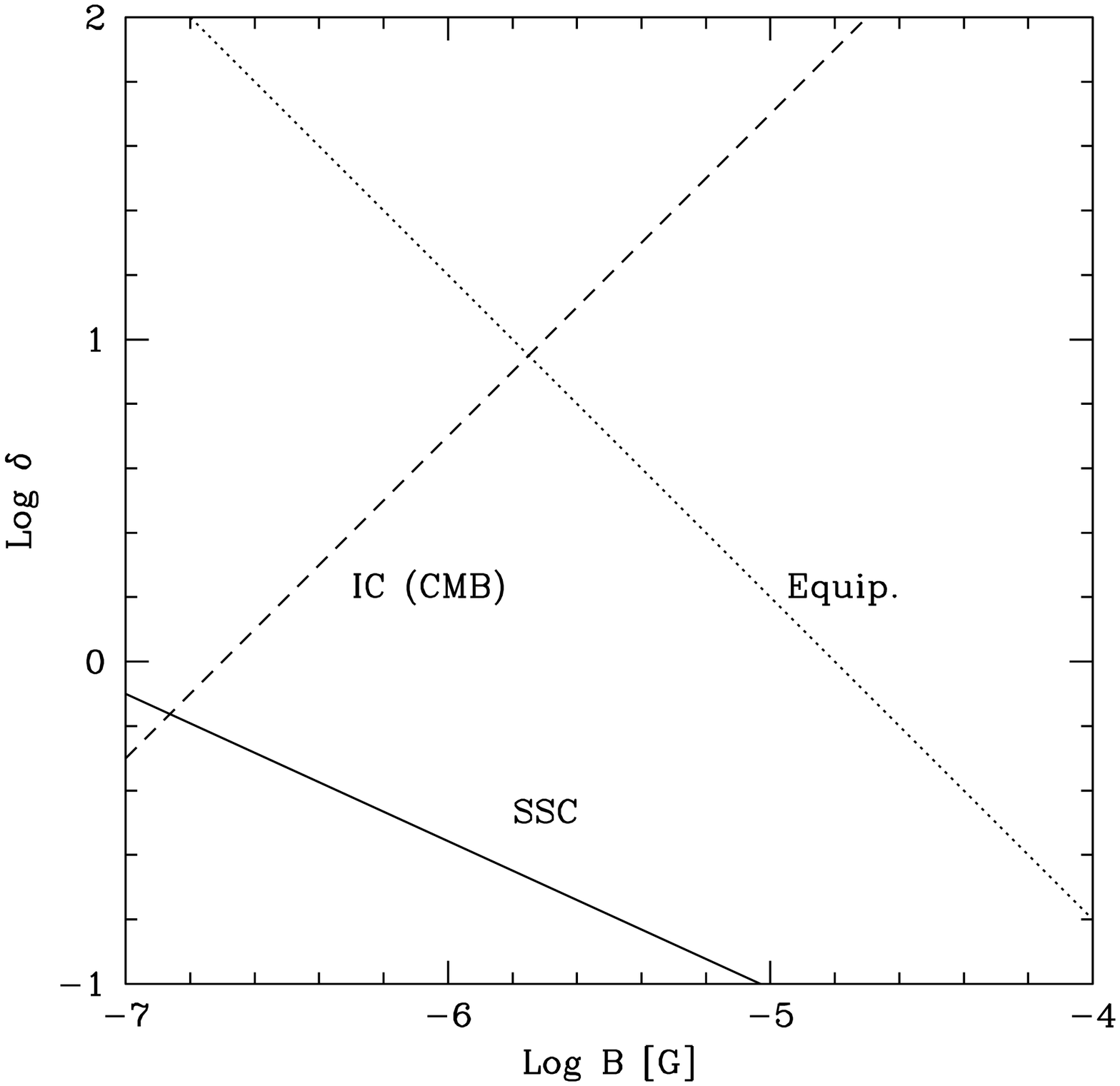,height=7.2cm,width=8.7cm}}  
\caption{Plot of the jet Doppler factor, $\delta$, versus the magnetic  
field, $B$, as expected from the IC/CMB process (dashed line), SSC
process (continuous line), and from the equipartition condition
(dotted line). Assuming equipartition, an X-ray origin via IC/CMB
implies beaming, $\delta \sim 10$. Weak beaming is possible provided
the magnetic field is very small, \ltsima $0.1\mu$Gauss, and out of
equipartition.}
\end{figure}    
To examine this in more detail, we plot in Fig. 6 the relationship  
between the Doppler factor, $\delta$, and the magnetic field, $B$, for  
the IC/CMB process (dashed line), the SSC process (continuous line),  
and the equipartition condition (dotted line). Assuming equipartition,  
the IC/CMB process accounts for the X-ray emission of the outer jet if  
$\delta \sim 10$ and $B \sim B_{\rm eq} \sim 2~\mu$Gauss.  Unless  
the magnetic field is extremely small ($<< 0.1~\mu$Gauss), SSC  
requires de-beaming.   
  
Assuming the equipartition magnetic field, the radiative time of the  
synchrotron electrons at 1.4 GHz is $t_{\rm r} \sim 10^{8}$ yr. This  
is much larger than the light crossing time of the radio/X-ray  
emission region, $t_{\rm cross} \sim 2.4 \times 10^4$ yr, accounting  
for the diffuse morphology at radio and X-rays.   
  
The assumption of IC/CMB in equipartition requires a strongly beamed 
($\delta \sim 10$) jet at 3.5\arcmin\ (so at least 93.7 kpc) from the 
core. For a jet inclination \gtsima 30\deg, as expected from unifying 
schemes (e.g., Urry \& Padovani 1995), this implies large Lorentz 
factors for the emitting plasma. Alternatively, weaker beaming 
($\delta << 10$) is allowed for a magnetic field at least one order of 
magnitude away from equipartition (Fig. 6). 
  
We thus explored other possibilities accounting for the concave 
radio-to-X-rays spectrum in jets, such as the model proposed by Dermer 
\& Atoyan (2002). The latter authors interpret the continuum from 
radio to X-rays as synchrotron emission from a {\it single electron 
distribution}, whose shape, due to the different IC cooling rate 
suffered by electrons with different energies, produces the optical 
``valley''. The model assumes that the electron cooling is dominated 
by IC scattering of the CMB photons. Low-energy electrons scatter the 
CMB in the Thomson limit, while the scattering by high-energy 
electrons, with Lorentz factor above $\gamma_{\rm KN}\sim mc^2/h \nu 
^{\prime}_{\rm CMB}$, will be suppressed by the Klein-Nishina decline 
of the cross section. Due to the low cooling rate of the high-energy 
electrons, the resulting electron energy distribution will be 
characterized by a flattening, at large Lorentz factors, which in turn 
produces a ``bump'' in the $\nu F(\nu)$ synchrotron spectrum. 
\begin{figure}  
\centerline{\psfig{figure=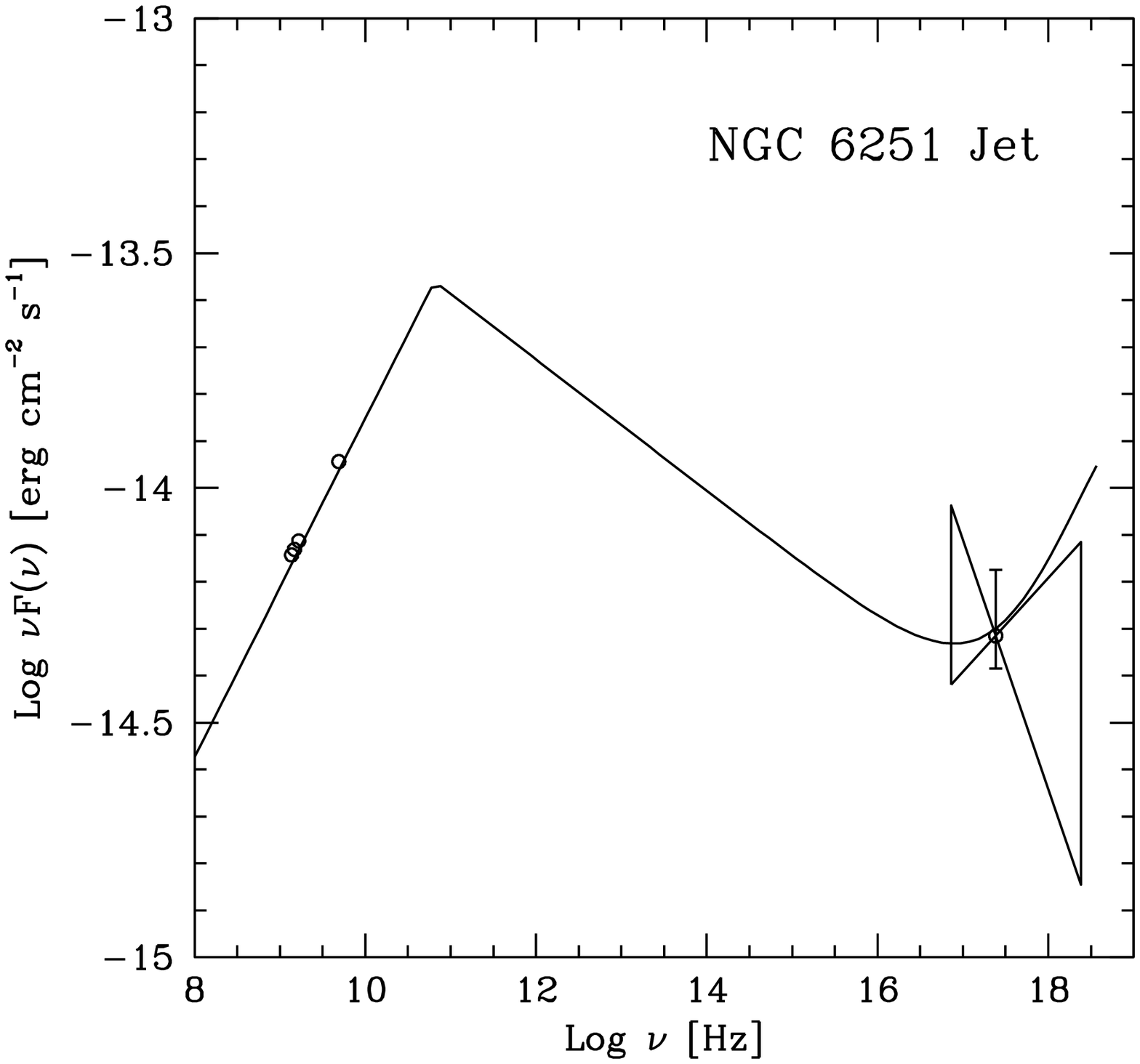,height=7.2cm,width=8.7cm}}  
\caption{Fit to the radio-to-X-ray spectral energy distribution  
of the jet with the Dermer \& Atoyan (2002) model. We assumed that the 
Lorentz factor of cooled particles (marking the spectral break at 
$10^{11} $ Hz) is $\gamma_{\rm cool}=1.5 \times 10^5$. Particles are 
injected at a rate $Q=7\times 10^{49}$ sec$^{-1}$, with slope 
$p=2.28$.}  
\end{figure}
   
We have calculated the expected spectrum implementing the simple 
analytical formulas provided by Dermer \& Atoyan (2002). We have 
assumed that the emission is very weakly beamed ($\delta=1$). The 
result is presented in Fig. 7. Note that, due to the assumed 
dominance of IC losses, the magnetic field is constrained to be less 
than $B\sim (8\pi U_{\rm CMB})^{0.5} \sim 10^{-6}$ G (assuming no 
beaming). In our case this implies that the emitting plasma is 
strongly far from equipartition, since the number of electrons required 
to produce the observed flux implies $U_{\rm e}\sim 10^{-9}$ erg 
cm$^{-3}$ (weakly dependent on the minimum Lorentz factor $\gamma_{\rm 
min}$). 
 
To conclude, we cannot identify a simple viable way to account for the
observed X-ray emission from the jet, since all the possibilities
explored above present important difficulties. However, it seems
possible to conclude that the jet is not confined by the external
gas. In fact, assuming a halo size of 200\arcsec\ or $2.86 \times
10^{23}$ cm, the density of the gas is $n_e \sim 2 \times 10^{-3}$
cm$^{-3}$, implying a gas pressure of $P_{\rm ext} \sim 5 \times 10^{-12}$
erg cm$^{-3}$.  If the plasma in the jet is not too far from
equipartition with the magnetic field (see dotted line in Fig. 6),
this value is smaller than or very close to the pressure inside the
jet. On the other hand, even if the magnetic field is well below
equipartition, the pressure of the relativistic electrons necessary to
account for the observed X-ray emission, both in the IC/CMB model and
in the Dermer \& Atoyan model, is much larger then the pressure in the
external gas. We conclude the X-ray jet is overpressured with respect
to the external medium, and cannot be confined by the halo, in
agreement with previous findings based on \rosat\ (Mack et al. 1997a,
Birkinshaw \& Worrall 1993).
  
\section{Summary}   
  
We presented a 40 ks \xmm\ EPIC observation of the giant elliptical  
\ngc, host of an FRI/II radio galaxy. An archival 25 ks \chandra\ 
exposure was also used. Our results can be summarized as follows: 
  
\begin{itemize}  
  
\item X-ray emission from the region of the 4\arcmin\ jet is detected, as well as  
diffuse emission from the galaxy's halo.  
  
\item The X-ray spectrum of the halo is described by a thermal model  
with temperature $kT \sim 1.6$ keV and sub-solar abundances, $Z \sim  
0.1 Z_{\odot}$.   
  
\item A hard X-ray component is required to fit the EPIC spectrum of  
the halo above 3 keV. This component is best described by a very flat,  
$\Gamma \sim 0$ power law, accounting for 20\% of the total 0.5--10  
keV flux. The hard component is also independently confirmed by the 
\chandra\ exposure.  
  
\item The physical origin of the hard X-ray tail in the halo is not 
clear. We argue that contribution from a population of faint Low Mass 
X-ray Binaries in unlikely. If due to IC/CMB, the weak diffuse radio 
emission and the X-ray flux imply a very low magnetic field, $<< 1 
\mu$Gauss, or a very narrow energy distribution of the high-energy 
electrons. Non-thermal bremsstrahlung from subrelativistic particles 
implies the presence of a large number of energetic electrons. 

\item The jet X-ray spectrum is described by a power law with photon  
index $\Gamma \sim 1.2$. The thermal model advocated on the basis of 
previous \rosat\ data (Mack et al. 1997a) is formally ruled out by the 
EPIC and ACIS data. However, based on the limited signal-to-noise 
ratio of both datasets, the possibility that a fraction of the jet 
X-ray emission is thermal remains open.  
  
\item An IC/CMB interpretation of the jet X-ray emission implies  
strong beaming ($\delta \sim 10$) at equipartition. Alternatively, 
weak beaming is possible if the magnetic field in the jet is more than 
one order of magnitude below equipartition.  A similar conclusion 
holds from the Dermer \& Atoyan (2002) model. 
\end{itemize}  
\vskip 0.4 cm  
\noindent{\centerline{{\bf Appendix: X-ray point sources in the Chandra field}  } 
\vskip 0.2 cm  
A glance at the \chandra\ image reveals the presence of several point  
sources in the field of the target. We used the \verb+CIAO+ tool  
{\tt wavdetect} to search for serendipitous X-ray sources in the  
field of view. In the algorithm, the parameter {\tt scale} (a list of radii,  
in image pixels, of Mexican Hat wavelet functions) was left free to  
range between 1 and 16, and the parameter {\tt threshold} (the number  
of detected spurious sources in a pixel map) held fixed at  
{$10^{-9}$}.  The algorithm returns a list of elliptical regions that  
define the positions and the shapes of the detected sources. We next  
used the coordinates from {\tt wavdetect} and its associated error  
regions to search for their optical counterparts on ESO archival  
plates. We used circular search regions with radii of 5\arcsec\  
because it is known that most reprocessed ACIS-I observations have an  
offset of up to 1.5\arcsec. \footnote[3]{See the memo on astrometry  
problems at http://cxc.\  
harvard.edu/\-mta/\-ASPECT/\-improve\_as\-trom\-e\-try.html}  
The properties of the point sources are listed in Table 1.   
\begin{table}[h]  
\caption{Field Sources}  
\begin{center}  
\begin{tabular}{cccccc}  
\hline
\hline  
\noalign{\smallskip}  
        R.A. &      DEC. &     X &     Y  &  Rate (c/s) & Notes \\  
\noalign{\smallskip}       
\hline			  	  	   	       
   248.09961 &  82.47877 &  4177 &  3596  $\pm$ 21.02 &    4.69 &   \\  
   247.97225 &  82.48873 &  4299 &  3669  $\pm$ 13.75 &    3.74 &  (a) \\ 
   247.80649 &  82.54228 &  4455 &  4062  $\pm$   5.83 &    2.45 &   \\  
   247.67471 &  82.53810 &  4580 &  4032  $\pm$  34.00 &    5.92 &   \\  
   248.36962 &  82.53544 &  3920 &  4011  $\pm$  4.80 &    2.24  &  \\  
   248.33852 &  82.53210 &  3949 &  3986  $\pm$ 16.65 &    4.12  &   \\  
   248.25310 &  82.52024 &  4030 &  3899  $\pm$  5.80 &    2.45  &  \\  
   248.23887 &  82.52276 &  4044 &  3918  $\pm$  4.81 &    2.24  &  \\  
   248.51997 &  82.51474 &  3776 &  3860  $\pm$  4.87 &    2.24  &  (b)\\  
   248.16867 &  82.57664 &  4111 &  4312  $\pm$  7.68 &    2.83  \\  
\hline  
\end{tabular}  
\end{center}  
{\bf Columns}: 1=Right Ascension at J2000; 2=Declination at J2000;   
3, 4=Physical Coordinates on the detector; 5=Count rate and error in the  
energy range 0.3--9 keV.\\  
(a) optically identified, U1650-01867603, RA=247.9735,  
DEC=+82.48894, $R$=17.5 mag,  $B$=19.0 mag, offset from X-ray source =0.016\arcmin.\\  
(b) optically identified, U1725-00533708 (NGC 6252), RA=248.16904,   
DEC=+82.57678, $R$=9.2 mag,  $B$=11.4 mag, offset from X-ray source =0.008\arcmin.\\  
Outside the extraction radius of 4\arcmin\ there are two more optically  
identified sources:\\  
(c1) U1650-01870981 , RA=248.63446,   
DEC=+82.48589, $R$=12.1 mag, $B$=12.6 mag;\\  
(c2) U1725-00532074 (2MASS J16290730+8231573), RA=247.27754,   
DEC=+82.53217, $R$=15.5 mag, $B$=16.6 mag.  
\end{table}  
\vskip 0.5 cm

\begin{acknowledgements}  
We thank the referee for useful suggestions which improved the 
discussion session.  RMS, MG, and DD are funded through NASA grants 
NAG5-10038, NAG5-10073, and LTSA NAG5-10708.  In addition, RMS 
gratefully acknowledges support from an NSF CAREER award and from the 
Clare Boothe Luce Program of the Henry Luce Foundation.  Radio 
astronomy at Brandeis University is supported by the NSF.  Further 
support to CCC by NASA grant GO2-3195C from the Smithsonian 
Observatory is gratefully acknowledged.    
\end{acknowledgements}

\end{document}